\documentclass[10pt,
 aip,
 jmp,%
 amsmath,amssymb,
 twocolumn,
reprint,%
]{revtex4-1}

\usepackage[linkcolor=blue]{hyperref}%
\usepackage{graphicx}%
\usepackage{graphics}%
\usepackage{epstopdf}
\usepackage{dcolumn}%
\usepackage{bm}% bold math
\usepackage[mathlines]{lineno}% Enable numbering of text and display math
\expandafter\ifx\csname package@font\endcsname\relax\else
 \expandafter\expandafter
 \expandafter\usepackage
 \expandafter\expandafter
 \expandafter{\csname package@font\endcsname}%
\fi
\begin{document}

\preprint{AIP/123-QED}

\title{Superconductivity in the system \texorpdfstring{$\rm Mo_xC_yGa_zO_{\delta}$}{Mo(x)C(y)Ga(z)O(delta)} prepared by focused ion beam induced deposition}
\author{P. M. Weirich}
		\email{p.weirich@Physik.uni-frankfurt.de}
    \affiliation{Physikalisches Institut, Goethe-University, 60438 Frankfurt am Main, Germany}
\author{C. H. Schwalb}
    \affiliation{Physikalisches Institut, Goethe-University, 60438 Frankfurt am Main, Germany}
\author{M. Winhold}
		\affiliation{Physikalisches Institut, Goethe-University, 60438 Frankfurt am Main, Germany}
\author{M.~Huth}
    \affiliation{Physikalisches Institut, Goethe-University, 60438 Frankfurt am Main, Germany}
\date{\today}

\begin{abstract}
We have prepared the new amorphous superconductor $\rm Mo_xC_yGa_zO_{\delta}$ with a maximum critical temperature $T_c$ of 3.8\,K by the direct-write nano-patterning technique of focused (gallium) ion beam induced deposition (FIBID) using $\rm Mo(CO)_6$ as precursor gas. From a detailed analysis of the temperature-dependent resistivity and the upper critical field we found clear evidence for proximity of the samples to a disorder-induced metal-insulator transition. We observed a strong dependence of $T_c$ on the deposition parameters and identified clear correlations between $T_c$, the localization tendency visible in the resistance data and the sample composition. By an in-situ feedback-controlled optimization process in the FIB-induced growth we were able to identify the beam parameters which lead to samples with the largest $T_c$-value and sharpest transition into the superconducting state.
\end{abstract}

%\keywords{}%Use showkeys class option if keyword
\maketitle

The capability for direct writing of superconducting nanostructures is attractive for a wide range of research fields and applications. Here we specifically mention the fabrication of structures for tunneling spectroscopy \cite{Guillamon2008} or for point contacts in Andreev junctions, e.g., with ferromagnetic counter electrodes \cite{Sangiao2011} and the use of sub-micron sized superconducting electrodes that induce exotic spin-triplet pairing in ferromagnetic nano wires \cite{Kompaniiets2014,Kompaniiets2014a}. So far the amorphous superconductor $\rm W_xC_yGa_zO_{\delta}$ prepared by focused ion beam induced deposition with a Ga source (Ga-FIBID) using the precursor $\rm W(CO)_6$ has been the only example in this regard \cite{Sadki2004}. However, in using this amorphous superconductor an important issue has not been adequately addressed, namely the high degree of structural disorder of this material which puts it close to a disorder-induced metal-insulator or Anderson transition \cite{Mott1975}. This has severe consequences with regard to various key quantities of the superconducting state, such as the dependence of the critical temperature, $T_c$, on the degree of disorder and the growing importance of statistical fluctuations of the superconducting order parameter leading to a spatial variation of the critical temperature $T_c(\textbf{r})$ \cite{Sadovskii1997}. In this case the usability of such a superconductor is severely limited. Here we introduce the system $\rm Mo_xC_yGa_zO_{\delta}$ prepared by Ga-FIBID and using $\rm Mo(CO)_6$ as a second example of a direct-write amorphous superconductor. We specifically address the issue of the interplay of disorder and interaction effects and show that this material is also close to an Anderson transition in three dimensions. We systematically study the influence of the dwell time (see below) as a very important deposition parameter on the sample composition, the apparent degree of disorder and the critical temperature. We show how optimized beam parameters can be readily found that lead to the highest critical temperature and, in parallel, to the most homogenous superconducting state. This is accomplished by monitoring the conductance of the deposited FIBID structures in-situ, coupled to a feedback loop that uses a genetic algorithm to change the deposition parameters as the growth proceeds such that the conductance of the samples is maximized \cite{Weirich2013}.

%%%%%%%%%%%%%%%%%%%%%%%%%%%%%%%%%%%%%%%%%%%%%%%%%%%%%%%%%%%%%%%
% Table 1
%%%%%%%%%%%%%%%%%%%%%%%%%%%%%%%%%%%%%%%%%%%%%%%%%%%%%%%%%%%%%%%

\begin{table*}
\caption{Summary of the FIBID parameters and properties of the samples. For samples 1-5 different dwell-times were applied, while the pitch was kept constant. For the deposition of sample 6, optimized parameters, obtained with the genetic algorithm, were utilized. For the deposition of each sample with a lateral dimension of 1x35\,$\rm \mu m^2$ a dose of $\rm 643\,pC/ \mu m^2$ was employed. For the determination of the resisitivities of the samples AFM and I(V)-measurements were performed. For the resistivities an uncertainty of 30\,$\%$ is assumed, due to FIB-induced etching in the early stage of the FIBID process.}
\label{table1}
\begin{ruledtabular}
\begin{tabular}{cccccccccccc}
 Sample & $t_D$ & \multicolumn{2}{c}{Pitch} & \multicolumn{4}{c}{Chemical composition} & $T_c$ & $B_{c2}$ & $\xi(0)$ & $\rho$\\
 &  & x & y & C & O & Ga & Mo & & & & \\ 
 \# & (ns) & \multicolumn{2}{c}{(nm)} & (at \%) & (at \%) & (at \%) & (at \%) & (K) & (T) & (nm) & $(\rm \mu \Omega cm)$ \\ \hline
 1 & 100 & 20 & 20 & 23.4 & 7.1 & 30.7 & 38.8 & 2.7 & 2.6 & 11.2 & $322.4 \pm 96.72$ \\
 2 & 250 & 20 & 20 & 22.6 & 11.7 & 29.7 & 36 & 2.9 & 2.5 & 7.6 & $464.3 \pm 133.29$ \\ 
 3 & 500 & 20 & 20 & 20.8 & 2.6 & 31.3 & 45.3 & 3.5 & 4.5 & 8.5 & $657 \pm 197.1$ \\ 
 4 & 750 & 20 & 20 & 23.3 & 8.5 & 30.4 & 37.8 & 3.3 & 3.7 & 9.4 & $616.2 \pm 184.86$ \\ 
 5 & 1000 & 20 & 20 & 23.8 & 8.1 & 30.6 & 37.5 & 2.9 & 3 & 10.5 & $518.9 \pm 155.67$ \\ 
 6 (GA)& 400 & 30 & 58 & 25.7 & 6.7 & 26.1 & 41.5 & 3.8 & 4.5 & 8.5 & $544.9 \pm 163.47$ \\ 
\end{tabular}
\end{ruledtabular}
\end{table*}

%%%%%%%%%%%%%%%%%%%%%%%%%%%%%%%%%%%%%%%%%%%%%%%%%%%%%%%%%%%%%%%

The FIBID process was performed in a dual-beam scanning electron microscope (SEM) (FEI, Nova Nanolab 600) equipped with a Ga-Ion source. Via a gas injection system the precursor gas was introduced into the high-vacuum chamber through a thin capillary with a diameter of 0.5\,mm in close proximity to the focus of the ion beam which was operated at a voltage of 30\,keV and a beam current of 10\,pA. For the deposition the temperature of the precursor $\rm Mo(CO)_6$ was set to $30^{\circ}$C. All samples were deposited on n-doped Si(100) substrates with a 300\,nm thick $\rm Si_3N_4$-layer. Employing standard UV lithography and lift-off the substrates were pre-patterned with Cr/Au contact structures of 10\,nm and 200\,nm thickness, respectively. The contact separation was 3\,$\mu$m. During the FIBID process the time-dependent conductance of the deposits was measured in-situ using a Keithley\,2400 sourcemeter. Samples 1 to 5 were obtained by varying the dwell time, t$_D$, i.e., the time the beam rests at a given position of the raster pattern that defines the FIBID-structure, for a fixed distance between the dwell points of 20\,nm (pitch). For sample 6 the deposition parameters dwell time and pitch in x- and y-direction were independently varied with a genetic algorithm that searched for the conditions leading to the fastest increase of the conductance as the growth proceeded (see below for details). After growth the averaged sample composition was determined by energy dispersive X-ray analysis (EDX) at a beam voltage of 5\,keV. Atomic force microscopy (AFM) measurements were done in dynamical mode (AFM $\rm Workshop_{TM}$ TT-AFM). Temperature-dependent resistance measurements were performed at a fixed bias current in a cryostat equipped with variable temperature insert and superconducting solenoid. Measurements of the upper critical field were done by temperature sweeps at fixed applied magnetic field. The onset temperature for the superconducting transition was defined as the temperature at which the extrapolated normal state resistance dropped by 10\,$\%$. In all of these measurements care was taken to keep the current density level below about 10\,A/cm$^2$ in order to prevent critical current induced shifts of the transition.

%%%%%%%%%%%%%%%%%%%%%%%%%%%%%%%%%%%%%%%%%%%%%%%%%%%%%%%%%%%%%%%
% Figure 1
%%%%%%%%%%%%%%%%%%%%%%%%%%%%%%%%%%%%%%%%%%%%%%%%%%%%%%%%%%%%%%%

\begin{figure}
\centering
    \includegraphics[width=0.48\textwidth]{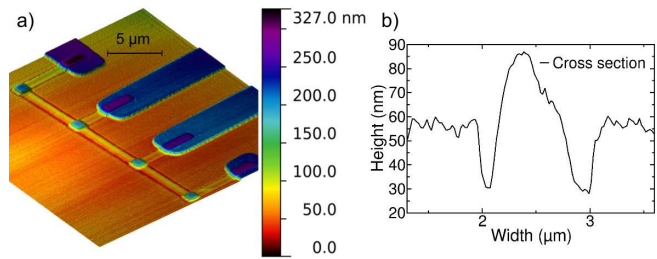}
    \caption[t]
    {a) An Exemplary AFM image of a four-probe $\rm Mo_xC_yGa_zO_{\delta}$ FIBID-structure used in this work. Indications for FIB-induced etching in the early stage of the growth process are apparent in b). b) Cross sections, as shown here for a sample deposited with a dwell-time of 250\,ns.}
   \label{figure1}
\end{figure}

%%%%%%%%%%%%%%%%%%%%%%%%%%%%%%%%%%%%%%%%%%%%%%%%%%%%%%%%%%%%%%%

\begin{figure*}
\centering
    \includegraphics[width=1.0\textwidth]{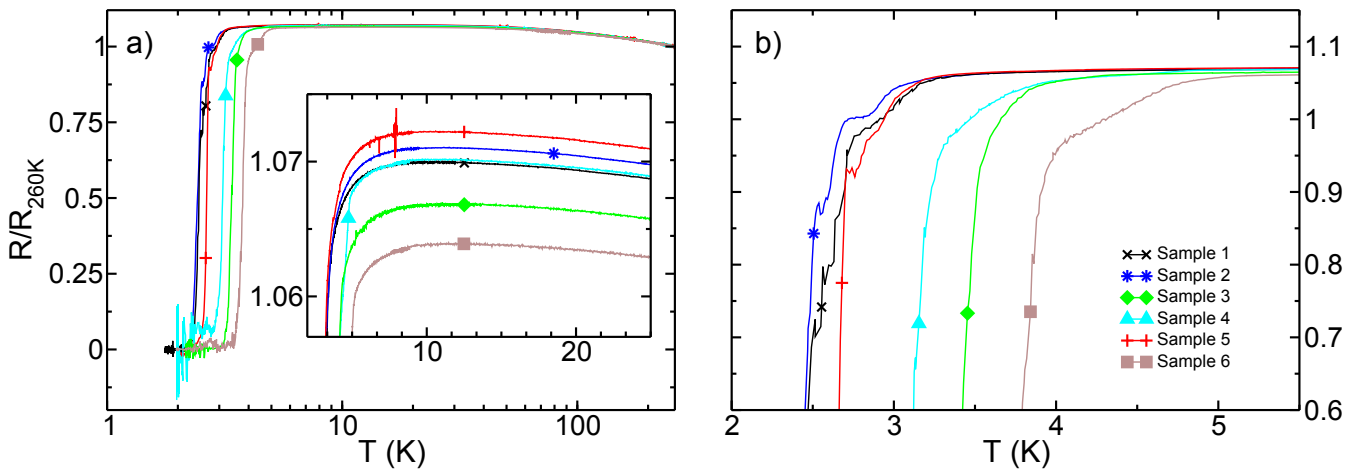}
    \caption[]
    {a) Temperature-dependent resistance of six samples, normalized to the respective resisitance values at 260\,K. A $T_c$ in the range from 2.7 to 3.8\,K is observed. Sample 6, deposited with optimized parameters ,shows the largest $T_c$. The inset shows a small negative temperature coefficient $\alpha=dR/dT$ for all samples be observed in the range from 10 to 25\,K. b) Detailed view on the temperature-dependent resistance, normalized to the value at 260\,K in the range from 2 to 5.5 \,K. The R(T)-curves exhibit a pronounced rounding far above the superconducting onset. For samples 1 and 2 multi-step or incomplete transitions occure. For sample 6 a rather distinct slope change is observed at 5\,K.}
   \label{figure2}
\end{figure*}

\begin{figure}
\centering
    \includegraphics[width=0.48\textwidth]{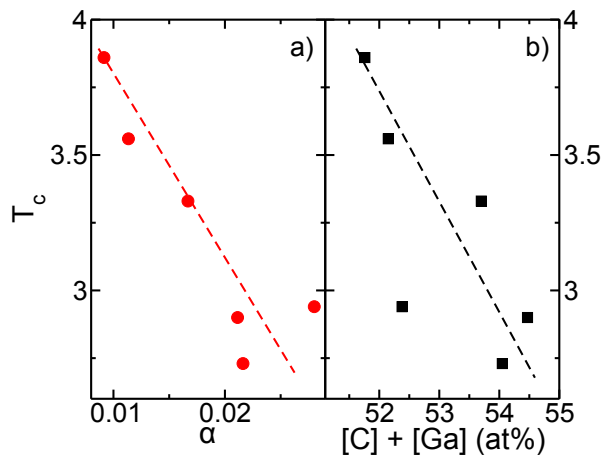}
    \caption[]
    {a) Plot of the superconducting transition against the small negative temperature coefficient $\alpha=dR/dT$ for all samples in the range from 10 to 25\,K showing a pronounced correlation . The dotted line serves to guide the eye. b) Plot of the superconducting transition temperature against the sum elemental contributions of C and Ga also showing a clear correlation.}
   \label{figure3}
\end{figure}

We begin the presentation of our results by showing in Fig. \ref{figure1} an exemplary AFM image of the four-probe $\rm Mo_xC_yGa_zO_{\delta}$ FIBID-structures used in this work. In the AFM line scan analysis we observed for all samples indications for FIB-induced etching in the early stages of the growth process, independent of the employed dwell times (see also Fig. \ref{figure1}). We speculate that the deposition rate in the early stage of growth is too small to compensate for the FIB-induced etching but increases as the process continues. As a consequence, the $\rm Mo_xC_yGa_zO_{\delta}$ structures are positioned in a well-like depression which prevented us from accurately determining the height of the deposits. This results in an uncertainty of about $30\,\%$ in the resistivity values calculated from the measured resistance values, as listed in Tab. \ref{table1}. Under variation of the dwell time from 100\,ns to 1000\,ns the measured resistance values at room temperature varied by a factor of up to 3.5. However, using the approximated sample heights, as determined by the AFM measurement, the resulting resistivity values varied by less than a factor of 2. For all samples this leads to resistivity values of 300 to 600\,$\mu\Omega$cm (see Tab. \ref{table1} for detailed numbers). This puts all the samples close to the critical resistivity for a disorder-induced metal-insulator transition \cite{Sadovskii1997}. Indeed, in Fig. \ref{figure2} we provide an overview graph of the temperature-dependent resistance, normalized to the value at 260\,K, that shows a small negative temperature coefficient $\alpha=dR/dT$ for all samples. We deduce the coefficients $\bar{\alpha}$ from the averaged slopes of the $R(T)$ curves taken in the range from 10 to 25\,K and plot in Fig.\ref{figure3}a the observed onset temperatures for the superconducting transition against these. We observe a pronounced correlation such that the larger the coefficient the smaller $T_c$. At this stage we may speculate that this is an indication for the close proximity to a disorder-induced metal-insulator transition for which the growth of Coulomb correlations leads to a reduction of $T_c$ \cite{Anderson1983}. Another indication of this proximity is the occurrence of multi-step or incomplete transitions which is evident from Fig. \ref{figure2} for samples 1 and 2. For these samples the self-averaging of the superconducting order parameter over the complete sample volume is not complete any more. Very close to the critical resistivity value statistical fluctuations of the order parameter can lead to a spatial inhomogeneity of the critical temperature $T_c(\textbf{r})$ which becomes apparent as appreciably broadened or multi-step transitions \cite{Sadovskii1997}. The $R(T)$-curves furthermore exhibit a pronounced rounding already far above the actual superconducting onset which we consider as a signature of the broadened critical region above $T_c$ in which fluctuational Cooper pair formation occurs and which is quite typical for disordered systems \cite{Sadovskii1997, Aslamazov1968}.

%%%%%%%%%%%%%%%%%%%%%%%%%%%%%%%%%%%%%%%%%%%%%%%%%%%%%%%%%%%%%%%
% Figure 2
%%%%%%%%%%%%%%%%%%%%%%%%%%%%%%%%%%%%%%%%%%%%%%%%%%%%%%%%%%%%%%%

%%%%%%%%%%%%%%%%%%%%%%%%%%%%%%%%%%%%%%%%%%%%%%%%%%%%%%%%%%%%%%%

In order to elucidate the influence of the sample composition on the properties of the samples we did EDX experiments and searched for possible correlations with the critical temperatures and the averaged temperature coefficients of resistance. We could identify one clear correlation between $T_c$ and the sum elemental contributions of C and Ga, such that the largest $T_c$ is found for the smallest Ga and C sum contribution, as is shown in Fig. \ref{figure3}b. On first glance this may suggest that $T_c$ grows with increasing metal content. However, we found this not to be the case, since also the O content varies with the dwell time. In order to shed more light on the question for which elemental composition the highest critical temperature and the most homogenous superconducting state can be obtained we now turn to the in-situ conductance procedure which we used.

%%%%%%%%%%%%%%%%%%%%%%%%%%%%%%%%%%%%%%%%%%%%%%%%%%%%%%%%%%%%%%%
% Figure 3
%%%%%%%%%%%%%%%%%%%%%%%%%%%%%%%%%%%%%%%%%%%%%%%%%%%%%%%%%%%%%%%

%%%%%%%%%%%%%%%%%%%%%%%%%%%%%%%%%%%%%%%%%%%%%%%%%%%%%%%%%%%%%%%

%%%%%%%%%%%%%%%%%%%%%%%%%%%%%%%%%%%%%%%%%%%%%%%%%%%%%%%%%%%%%%%
% Figure 4
%%%%%%%%%%%%%%%%%%%%%%%%%%%%%%%%%%%%%%%%%%%%%%%%%%%%%%%%%%%%%%%

\begin{figure}
\centering
    \includegraphics[width=0.48\textwidth]{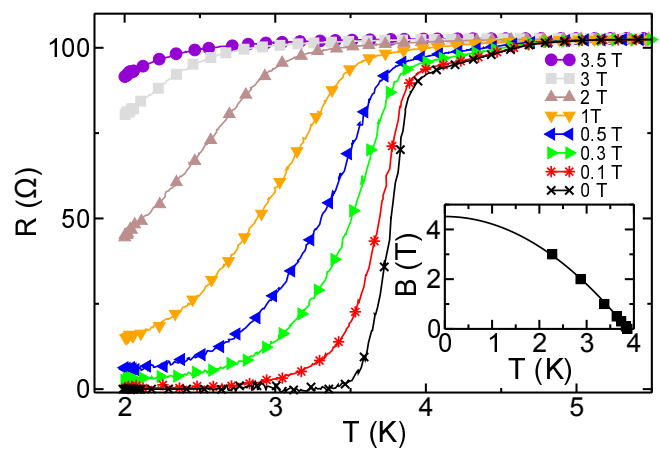}
    \caption[]
    {R(T)-curves for sample 6 in different magnetic fields. The field was applied perpendicular to the sample surface. The superconducting transition shows a homogenous broadening in parallel with a shift of the onset temperature as the field increases. Inset: Upper critical field values as obtained from R(T) curve onset temperatures. The solid line represents a Ginzburg-Landau fit, as detailed in the text.}
   \label{figure4}
\end{figure}

%%%%%%%%%%%%%%%%%%%%%%%%%%%%%%%%%%%%%%%%%%%%%%%%%%%%%%%%%%%%%%%

Very recently we have introduced a procedure for focused electron beam induced deposition (FEBID) that allows for a feedback-controlled identification of the deposition parameters which yield the largest (or smallest) conductance increase for a growing deposit \cite{Weirich2013}. In brief, the process uses conductance data acquired in-situ during the repeated rastering of the pattern with a fixed deposition parameter set (typically dwell time and pitch). For a given parameter set the respective rates of conductance increase are determined and a new parameter set is created using a genetic algorithm\cite{Haupt2004}. With this new parameter set the growth is continued and the conductance increase rates are again determined in a cyclic fashion. Details of this procedure can be found in \cite{Weirich2013}. In particular, for FEBID structures prepared with the precursor $\rm W(CO)_6$ we could show, that the optimized deposition parameter set does indeed lead to deposits with the largest conductivity and that parameter-dependent changes of the overall growth rate are rather insignificant \cite{Weirich2013,Winhold2014,Winhold2014a}. Based on these findings we employed this optimization scheme here for FIBID and identified the dwell time and pitch parameters that were used for sample 6. This sample showed the largest conductivity. As is shown in Fig. \ref{figure2}, sample 6 also has the largest critical temperature and shows a rather smooth transition. We analyzed this sample further and performed measurements of the upper critical field by temperature sweeps in constant magnetic field. As is shown in Fig. \ref{figure4}, the transition shows a homogenous broadening in parallel with a shift of the onset temperature as the field is increased. Using the Ginzburg-Landau relations $B_{c2}(T)=B_{c2}(0)(1-(T/T_c)^2)$ and $B_{c2}(0)=\Phi_0/2\pi\xi(0)^2$ ($\Phi$: flux quantum) we deduced the upper critical field at 0\,K and the coherence length $\xi(0)$. This same analysis was repeated for all samples and the thus obtained upper critical fields and coherence lengths are listed in Tab. \ref{table1}.

We conclude the analysis by another interesting observation. In the $R(T)$-curve for sample 6 a rather distinct change of slope occurs at about 5\,K, clearly far above the actual onset of the superconducting transition at 3.8\,K. We note, that the superconductor $\rm W_xC_yGa_zO_{\delta}$ typically shows the onset to superconductivity at 5\,K \cite{Sadki2004, Sadki2005, Li2008, Luxmoore2007, Ross2006}. From this observation one may speculate that the slope change in sample 6 is an indication for the formation of a non-percolative and inhomogenous superconducting phase with a critical temperature of 5\,K. The question then arises whether the specific properties of the refractory metal component, W or Mo, is really decisive for the formation of the superconducting state or if the substrate of the superconduting phase is rather the carbon component, as was recently speculated for W-based FEBID structures with sulphur doping \cite{Felner2012,Felner2013}. This has to be adressed in future studies.

%\bibliography{./APL_prox}
\bibliographystyle{aipnum4-1}
\bibliographystyle{apsrev4-1}

\bibliography{APLMoIBID2014}

\begin{thebibliography}{19}%
\makeatletter
\providecommand \@ifxundefined [1]{%
 \@ifx{#1\undefined}
}%
\providecommand \@ifnum [1]{%
 \ifnum #1\expandafter \@firstoftwo
 \else \expandafter \@secondoftwo
 \fi
}%
\providecommand \@ifx [1]{%
 \ifx #1\expandafter \@firstoftwo
 \else \expandafter \@secondoftwo
 \fi
}%
\providecommand \natexlab [1]{#1}%
\providecommand \enquote  [1]{``#1''}%
\providecommand \bibnamefont  [1]{#1}%
\providecommand \bibfnamefont [1]{#1}%
\providecommand \citenamefont [1]{#1}%
\providecommand \href@noop [0]{\@secondoftwo}%
\providecommand \href [0]{\begingroup \@sanitize@url \@href}%
\providecommand \@href[1]{\@@startlink{#1}\@@href}%
\providecommand \@@href[1]{\endgroup#1\@@endlink}%
\providecommand \@sanitize@url [0]{\catcode `\\12\catcode `\$12\catcode
  `\&12\catcode `\#12\catcode `\^12\catcode `\_12\catcode `\%12\relax}%
\providecommand \@@startlink[1]{}%
\providecommand \@@endlink[0]{}%
\providecommand \url  [0]{\begingroup\@sanitize@url \@url }%
\providecommand \@url [1]{\endgroup\@href {#1}{\urlprefix }}%
\providecommand \urlprefix  [0]{URL }%
\providecommand \Eprint [0]{\href }%
\providecommand \doibase [0]{http://dx.doi.org/}%
\providecommand \selectlanguage [0]{\@gobble}%
\providecommand \bibinfo  [0]{\@secondoftwo}%
\providecommand \bibfield  [0]{\@secondoftwo}%
\providecommand \translation [1]{[#1]}%
\providecommand \BibitemOpen [0]{}%
\providecommand \bibitemStop [0]{}%
\providecommand \bibitemNoStop [0]{.\EOS\space}%
\providecommand \EOS [0]{\spacefactor3000\relax}%
\providecommand \BibitemShut  [1]{\csname bibitem#1\endcsname}%
\let\auto@bib@innerbib\@empty
%</preamble>
\bibitem [{\citenamefont {Guillam\'on}\ \emph {et~al.}(2008)\citenamefont
  {Guillam\'on}, \citenamefont {Suderow}, \citenamefont {Vieira}, \citenamefont
  {Fern\'andez-Pacheco}, \citenamefont {Ses\'e}, \citenamefont {C\'odoba},
  \citenamefont {Teresa},\ and\ \citenamefont {Ibarra}}]{Guillamon2008}%
  \BibitemOpen
  \bibfield  {author} {\bibinfo {author} {\bibfnamefont {I.}~\bibnamefont
  {Guillam\'on}}, \bibinfo {author} {\bibfnamefont {H.}~\bibnamefont
  {Suderow}}, \bibinfo {author} {\bibfnamefont {S.}~\bibnamefont {Vieira}},
  \bibinfo {author} {\bibfnamefont {A.}~\bibnamefont {Fern\'andez-Pacheco}},
  \bibinfo {author} {\bibfnamefont {J.}~\bibnamefont {Ses\'e}}, \bibinfo
  {author} {\bibfnamefont {R.}~\bibnamefont {C\'odoba}}, \bibinfo {author}
  {\bibfnamefont {J.~M.~D.}\ \bibnamefont {Teresa}}, \ and\ \bibinfo {author}
  {\bibfnamefont {M.~R.}\ \bibnamefont {Ibarra}},\ }\href
  {http://stacks.iop.org/1367-2630/10/i=9/a=093005} {\bibfield  {journal}
  {\bibinfo  {journal} {New Journal of Physics}\ }\textbf {\bibinfo {volume}
  {10}},\ \bibinfo {pages} {093005} (\bibinfo {year} {2008})}\BibitemShut
  {NoStop}%
\bibitem [{\citenamefont {Sangiao}\ \emph {et~al.}(2011)\citenamefont
  {Sangiao}, \citenamefont {De~Teresa}, \citenamefont {Ibarra}, \citenamefont
  {Guillam\'on}, \citenamefont {Suderow}, \citenamefont {Vieira},\ and\
  \citenamefont {Morell\'on}}]{Sangiao2011}%
  \BibitemOpen
  \bibfield  {author} {\bibinfo {author} {\bibfnamefont {S.}~\bibnamefont
  {Sangiao}}, \bibinfo {author} {\bibfnamefont {J.~M.}\ \bibnamefont
  {De~Teresa}}, \bibinfo {author} {\bibfnamefont {M.~R.}\ \bibnamefont
  {Ibarra}}, \bibinfo {author} {\bibfnamefont {I.}~\bibnamefont {Guillam\'on}},
  \bibinfo {author} {\bibfnamefont {H.}~\bibnamefont {Suderow}}, \bibinfo
  {author} {\bibfnamefont {S.}~\bibnamefont {Vieira}}, \ and\ \bibinfo {author}
  {\bibfnamefont {L.}~\bibnamefont {Morell\'on}},\ }\href {\doibase
  10.1103/PhysRevB.84.233402} {\bibfield  {journal} {\bibinfo  {journal} {Phys.
  Rev. B}\ }\textbf {\bibinfo {volume} {84}},\ \bibinfo {pages} {233402}
  (\bibinfo {year} {2011})}\BibitemShut {NoStop}%
\bibitem [{\citenamefont {Kompaniiets}\ \emph {et~al.}(2014)\citenamefont
  {Kompaniiets}, \citenamefont {Dobrovolskiy}, \citenamefont {Neetzel},
  \citenamefont {Porrati}, \citenamefont {Br\"otz}, \citenamefont {Ensinger},\
  and\ \citenamefont {Huth}}]{Kompaniiets2014}%
  \BibitemOpen
  \bibfield  {author} {\bibinfo {author} {\bibfnamefont {M.}~\bibnamefont
  {Kompaniiets}}, \bibinfo {author} {\bibfnamefont {O.~V.}\ \bibnamefont
  {Dobrovolskiy}}, \bibinfo {author} {\bibfnamefont {C.}~\bibnamefont
  {Neetzel}}, \bibinfo {author} {\bibfnamefont {F.}~\bibnamefont {Porrati}},
  \bibinfo {author} {\bibfnamefont {J.}~\bibnamefont {Br\"otz}}, \bibinfo
  {author} {\bibfnamefont {W.}~\bibnamefont {Ensinger}}, \ and\ \bibinfo
  {author} {\bibfnamefont {M.}~\bibnamefont {Huth}},\ }\href {\doibase
  http://dx.doi.org/10.1063/1.4863980} {\bibfield  {journal} {\bibinfo
  {journal} {Applied Physics Letters}\ }\textbf {\bibinfo {volume} {104}},\
  \bibinfo {eid} {052603} (\bibinfo {year} {2014})}\BibitemShut {NoStop}%
\bibitem [{\citenamefont {Kompaniiets}\ \emph {et~al.}(2013)\citenamefont
  {Kompaniiets}, \citenamefont {Dobrovolskiy}, \citenamefont {Neetzel},
  \citenamefont {Begun}, \citenamefont {Porrati}, \citenamefont {Ensinger},\
  and\ \citenamefont {Huth}}]{Kompaniiets2014a}%
  \BibitemOpen
  \bibfield  {author} {\bibinfo {author} {\bibfnamefont {M.}~\bibnamefont
  {Kompaniiets}}, \bibinfo {author} {\bibfnamefont {O.~V.}\ \bibnamefont
  {Dobrovolskiy}}, \bibinfo {author} {\bibfnamefont {C.}~\bibnamefont
  {Neetzel}}, \bibinfo {author} {\bibfnamefont {E.}~\bibnamefont {Begun}},
  \bibinfo {author} {\bibfnamefont {F.}~\bibnamefont {Porrati}}, \bibinfo
  {author} {\bibfnamefont {W.}~\bibnamefont {Ensinger}}, \ and\ \bibinfo
  {author} {\bibfnamefont {M.}~\bibnamefont {Huth}},\ }\href
  {http://arxiv.org/abs/1310.6595} {\bibfield  {journal} {\bibinfo  {journal}
  {ArXiv e-prints}\ } (\bibinfo {year} {2013})},\ \Eprint
  {http://arxiv.org/abs/1310.6595} {arXiv:1310.6595 [cond-mat.supr-con]}
  \BibitemShut {NoStop}%
\bibitem [{\citenamefont {Sadki}\ \emph {et~al.}(2004)\citenamefont {Sadki},
  \citenamefont {Ooi},\ and\ \citenamefont {Hirata}}]{Sadki2004}%
  \BibitemOpen
  \bibfield  {author} {\bibinfo {author} {\bibfnamefont {E.~S.}\ \bibnamefont
  {Sadki}}, \bibinfo {author} {\bibfnamefont {S.}~\bibnamefont {Ooi}}, \ and\
  \bibinfo {author} {\bibfnamefont {K.}~\bibnamefont {Hirata}},\ }\href
  {\doibase http://dx.doi.org/10.1063/1.1842367} {\bibfield  {journal}
  {\bibinfo  {journal} {Applied Physics Letters}\ }\textbf {\bibinfo {volume}
  {85}},\ \bibinfo {pages} {6206} (\bibinfo {year} {2004})}\BibitemShut
  {NoStop}%
\bibitem [{\citenamefont {Mott}\ \emph {et~al.}(1975)\citenamefont {Mott},
  \citenamefont {Pepper}, \citenamefont {Pollitt}, \citenamefont {Wallis},\
  and\ \citenamefont {Adkins}}]{Mott1975}%
  \BibitemOpen
  \bibfield  {author} {\bibinfo {author} {\bibfnamefont {N.}~\bibnamefont
  {Mott}}, \bibinfo {author} {\bibfnamefont {M.}~\bibnamefont {Pepper}},
  \bibinfo {author} {\bibfnamefont {S.}~\bibnamefont {Pollitt}}, \bibinfo
  {author} {\bibfnamefont {R.~H.}\ \bibnamefont {Wallis}}, \ and\ \bibinfo
  {author} {\bibfnamefont {C.~J.}\ \bibnamefont {Adkins}},\ }\href {\doibase
  10.1098/rspa.1975.0131} {\bibfield  {journal} {\bibinfo  {journal}
  {Proceedings of the Royal Society of London. A. Mathematical and Physical
  Sciences}\ }\textbf {\bibinfo {volume} {345}},\ \bibinfo {pages} {169}
  (\bibinfo {year} {1975})}\BibitemShut {NoStop}%
\bibitem [{\citenamefont {Sadovskii}(1997)}]{Sadovskii1997}%
  \BibitemOpen
  \bibfield  {author} {\bibinfo {author} {\bibfnamefont {M.~V.}\ \bibnamefont
  {Sadovskii}},\ }\href {\doibase
  http://dx.doi.org/10.1016/S0370-1573(96)00036-1} {\bibfield  {journal}
  {\bibinfo  {journal} {Physics Reports}\ }\textbf {\bibinfo {volume} {282}},\
  \bibinfo {pages} {225 } (\bibinfo {year} {1997})}\BibitemShut {NoStop}%
\bibitem [{\citenamefont {Weirich}\ \emph {et~al.}(2013)\citenamefont
  {Weirich}, \citenamefont {Winhold}, \citenamefont {Schwalb},\ and\
  \citenamefont {Huth}}]{Weirich2013}%
  \BibitemOpen
  \bibfield  {author} {\bibinfo {author} {\bibfnamefont {P.~M.}\ \bibnamefont
  {Weirich}}, \bibinfo {author} {\bibfnamefont {M.}~\bibnamefont {Winhold}},
  \bibinfo {author} {\bibfnamefont {C.~H.}\ \bibnamefont {Schwalb}}, \ and\
  \bibinfo {author} {\bibfnamefont {M.}~\bibnamefont {Huth}},\ }\href {\doibase
  10.3762/bjnano.4.103} {\bibfield  {journal} {\bibinfo  {journal} {Beilstein
  Journal of Nanotechnology}\ }\textbf {\bibinfo {volume} {4}},\ \bibinfo
  {pages} {919} (\bibinfo {year} {2013})}\BibitemShut {NoStop}%
\bibitem [{\citenamefont {Anderson}\ \emph {et~al.}(1983)\citenamefont
  {Anderson}, \citenamefont {Muttalib},\ and\ \citenamefont
  {Ramakrishnan}}]{Anderson1983}%
  \BibitemOpen
  \bibfield  {author} {\bibinfo {author} {\bibfnamefont {P.~W.}\ \bibnamefont
  {Anderson}}, \bibinfo {author} {\bibfnamefont {K.~A.}\ \bibnamefont
  {Muttalib}}, \ and\ \bibinfo {author} {\bibfnamefont {T.~V.}\ \bibnamefont
  {Ramakrishnan}},\ }\href {\doibase 10.1103/PhysRevB.28.117} {\bibfield
  {journal} {\bibinfo  {journal} {Phys. Rev. B}\ }\textbf {\bibinfo {volume}
  {28}},\ \bibinfo {pages} {117} (\bibinfo {year} {1983})}\BibitemShut
  {NoStop}%
\bibitem [{\citenamefont {Aslamazov}\ and\ \citenamefont
  {Larkin}(1968)}]{Aslamazov1968}%
  \BibitemOpen
  \bibfield  {author} {\bibinfo {author} {\bibfnamefont {L.}~\bibnamefont
  {Aslamazov}}\ and\ \bibinfo {author} {\bibfnamefont {A.}~\bibnamefont
  {Larkin}},\ }\href@noop {} {\bibfield  {journal} {\bibinfo  {journal} {Soviet
  Physics - Solid State}\ ,\ \bibinfo {pages} {1104}} (\bibinfo {year}
  {1968})}\BibitemShut {NoStop}%
\bibitem [{\citenamefont {Haupt}\ and\ \citenamefont
  {Haupt}(2004)}]{Haupt2004}%
  \BibitemOpen
  \bibfield  {author} {\bibinfo {author} {\bibfnamefont {R.~L.}\ \bibnamefont
  {Haupt}}\ and\ \bibinfo {author} {\bibfnamefont {S.~E.}\ \bibnamefont
  {Haupt}},\ }\href {http://books.google.de/books?id=5gR97w4X0EkC} {\emph
  {\bibinfo {title} {Practical Genetic Algorithms}}}\ (\bibinfo  {publisher}
  {Wiley},\ \bibinfo {year} {2004})\BibitemShut {NoStop}%
\bibitem [{\citenamefont {Winhold}\ \emph
  {et~al.}(2014{\natexlab{a}})\citenamefont {Winhold}, \citenamefont {Weirich},
  \citenamefont {Schwalb},\ and\ \citenamefont {Huth}}]{Winhold2014}%
  \BibitemOpen
  \bibfield  {author} {\bibinfo {author} {\bibfnamefont {M.}~\bibnamefont
  {Winhold}}, \bibinfo {author} {\bibfnamefont {P.~M.}\ \bibnamefont
  {Weirich}}, \bibinfo {author} {\bibfnamefont {C.~H.}\ \bibnamefont
  {Schwalb}}, \ and\ \bibinfo {author} {\bibfnamefont {M.}~\bibnamefont
  {Huth}},\ }\href@noop {} {\enquote {\bibinfo {title} {Modeling the genetic
  algorithm optimization process in focused electron beam induced
  deposition},}\ } (\bibinfo {year} {2014}{\natexlab{a}}),\ \bibinfo {note}
  {submitted to Nanofabrication}\BibitemShut {NoStop}%
\bibitem [{\citenamefont {Winhold}\ \emph
  {et~al.}(2014{\natexlab{b}})\citenamefont {Winhold}, \citenamefont {Weirich},
  \citenamefont {Schwalb},\ and\ \citenamefont {Huth}}]{Winhold2014a}%
  \BibitemOpen
  \bibfield  {author} {\bibinfo {author} {\bibfnamefont {M.}~\bibnamefont
  {Winhold}}, \bibinfo {author} {\bibfnamefont {P.~M.}\ \bibnamefont
  {Weirich}}, \bibinfo {author} {\bibfnamefont {C.~H.}\ \bibnamefont
  {Schwalb}}, \ and\ \bibinfo {author} {\bibfnamefont {M.}~\bibnamefont
  {Huth}},\ }\href@noop {} {\enquote {\bibinfo {title} {Identifying the
  crossover between growth regimes via in-situ conductance measurements in
  focused electron beam induced deposition},}\ } (\bibinfo {year}
  {2014}{\natexlab{b}}),\ \bibinfo {note} {submitted to Microlelectronic
  Engineering}\BibitemShut {NoStop}%
\bibitem [{\citenamefont {Sadki}\ \emph {et~al.}(2005)\citenamefont {Sadki},
  \citenamefont {Ooi},\ and\ \citenamefont {Hirata}}]{Sadki2005}%
  \BibitemOpen
  \bibfield  {author} {\bibinfo {author} {\bibfnamefont {E.}~\bibnamefont
  {Sadki}}, \bibinfo {author} {\bibfnamefont {S.}~\bibnamefont {Ooi}}, \ and\
  \bibinfo {author} {\bibfnamefont {K.}~\bibnamefont {Hirata}},\ }\href
  {\doibase http://dx.doi.org/10.1016/j.physc.2005.02.151} {\bibfield
  {journal} {\bibinfo  {journal} {Physica C: Superconductivity}\ }\textbf
  {\bibinfo {volume} {426-431, Part 2}},\ \bibinfo {pages} {1547 } (\bibinfo
  {year} {2005})},\ \bibinfo {note} {proceedings of the 17th International
  Symposium on Superconductivity (ISS 2004) Advances in Superconductivity
  \{XVII\} Proceedings of the 17th International Symposium on Superconductivity
  (ISS 2004)}\BibitemShut {NoStop}%
\bibitem [{\citenamefont {Li}\ \emph {et~al.}(2008)\citenamefont {Li},
  \citenamefont {Fenton}, \citenamefont {Wang}, \citenamefont {McComb},\ and\
  \citenamefont {Warburton}}]{Li2008}%
  \BibitemOpen
  \bibfield  {author} {\bibinfo {author} {\bibfnamefont {W.}~\bibnamefont
  {Li}}, \bibinfo {author} {\bibfnamefont {J.~C.}\ \bibnamefont {Fenton}},
  \bibinfo {author} {\bibfnamefont {Y.}~\bibnamefont {Wang}}, \bibinfo {author}
  {\bibfnamefont {D.~W.}\ \bibnamefont {McComb}}, \ and\ \bibinfo {author}
  {\bibfnamefont {P.~A.}\ \bibnamefont {Warburton}},\ }\href {\doibase
  http://dx.doi.org/10.1063/1.3013444} {\bibfield  {journal} {\bibinfo
  {journal} {Journal of Applied Physics}\ }\textbf {\bibinfo {volume} {104}},\
  \bibinfo {eid} {093913} (\bibinfo {year} {2008})}\BibitemShut {NoStop}%
\bibitem [{\citenamefont {Luxmoore}\ \emph {et~al.}(2007)\citenamefont
  {Luxmoore}, \citenamefont {Ross}, \citenamefont {Cullis}, \citenamefont
  {Fry}, \citenamefont {Orr}, \citenamefont {Buckle},\ and\ \citenamefont
  {Jefferson}}]{Luxmoore2007}%
  \BibitemOpen
  \bibfield  {author} {\bibinfo {author} {\bibfnamefont {I.}~\bibnamefont
  {Luxmoore}}, \bibinfo {author} {\bibfnamefont {I.}~\bibnamefont {Ross}},
  \bibinfo {author} {\bibfnamefont {A.}~\bibnamefont {Cullis}}, \bibinfo
  {author} {\bibfnamefont {P.}~\bibnamefont {Fry}}, \bibinfo {author}
  {\bibfnamefont {J.}~\bibnamefont {Orr}}, \bibinfo {author} {\bibfnamefont
  {P.}~\bibnamefont {Buckle}}, \ and\ \bibinfo {author} {\bibfnamefont
  {J.}~\bibnamefont {Jefferson}},\ }\href {\doibase
  http://dx.doi.org/10.1016/j.tsf.2007.02.029} {\bibfield  {journal} {\bibinfo
  {journal} {Thin Solid Films}\ }\textbf {\bibinfo {volume} {515}},\ \bibinfo
  {pages} {6791 } (\bibinfo {year} {2007})}\BibitemShut {NoStop}%
\bibitem [{\citenamefont {Ross}\ \emph {et~al.}(2006)\citenamefont {Ross},
  \citenamefont {Luxmoore}, \citenamefont {Cullis}, \citenamefont {Orr},
  \citenamefont {Buckle},\ and\ \citenamefont {Jefferson}}]{Ross2006}%
  \BibitemOpen
  \bibfield  {author} {\bibinfo {author} {\bibfnamefont {I.~M.}\ \bibnamefont
  {Ross}}, \bibinfo {author} {\bibfnamefont {I.~J.}\ \bibnamefont {Luxmoore}},
  \bibinfo {author} {\bibfnamefont {A.~G.}\ \bibnamefont {Cullis}}, \bibinfo
  {author} {\bibfnamefont {J.}~\bibnamefont {Orr}}, \bibinfo {author}
  {\bibfnamefont {P.~D.}\ \bibnamefont {Buckle}}, \ and\ \bibinfo {author}
  {\bibfnamefont {J.~H.}\ \bibnamefont {Jefferson}},\ }\href
  {http://stacks.iop.org/1742-6596/26/i=1/a=088} {\bibfield  {journal}
  {\bibinfo  {journal} {Journal of Physics: Conference Series}\ }\textbf
  {\bibinfo {volume} {26}},\ \bibinfo {pages} {363} (\bibinfo {year}
  {2006})}\BibitemShut {NoStop}%
\bibitem [{\citenamefont {Felner}\ \emph {et~al.}(2012)\citenamefont {Felner},
  \citenamefont {Wolf},\ and\ \citenamefont {Millo}}]{Felner2012}%
  \BibitemOpen
  \bibfield  {author} {\bibinfo {author} {\bibfnamefont {I.}~\bibnamefont
  {Felner}}, \bibinfo {author} {\bibfnamefont {O.}~\bibnamefont {Wolf}}, \ and\
  \bibinfo {author} {\bibfnamefont {O.}~\bibnamefont {Millo}},\ }\href
  {\doibase 10.1007/s10948-011-1327-x} {\bibfield  {journal} {\bibinfo
  {journal} {Journal of Superconductivity and Novel Magnetism}\ }\textbf
  {\bibinfo {volume} {25}},\ \bibinfo {pages} {7} (\bibinfo {year}
  {2012})}\BibitemShut {NoStop}%
\bibitem [{\citenamefont {Felner}\ \emph {et~al.}(2013)\citenamefont {Felner},
  \citenamefont {Wolf},\ and\ \citenamefont {Millo}}]{Felner2013}%
  \BibitemOpen
  \bibfield  {author} {\bibinfo {author} {\bibfnamefont {I.}~\bibnamefont
  {Felner}}, \bibinfo {author} {\bibfnamefont {O.}~\bibnamefont {Wolf}}, \ and\
  \bibinfo {author} {\bibfnamefont {O.}~\bibnamefont {Millo}},\ }\href
  {\doibase 10.1007/s10948-013-2105-8} {\bibfield  {journal} {\bibinfo
  {journal} {Journal of Superconductivity and Novel Magnetism}\ }\textbf
  {\bibinfo {volume} {26}},\ \bibinfo {pages} {511} (\bibinfo {year}
  {2013})}\BibitemShut {NoStop}%
\end{thebibliography}%

\end{document}